\documentclass[11pt,a4paper]{article}
\usepackage{epsfig}
\usepackage{amsmath}

\newcommand{\avk}{\left< k \right>}
\newcommand{\fluck}{\left< k^2 \right>}
\newcommand{\condP}{P(k' \, \vert \, k)}
\newcommand{\cprob}{P^{nc}(k' \, \vert \, k)}
\newcommand{\condPN}[2]{P({#1} \, \vert \, {#2})}
\newcommand{\Ri}{R_\infty}
\newcommand{\Fi}{\phi_\infty}
\newcommand{\equ}[1]{(\protect\ref{#1})}

\begin{document}

\title{Epidemic spreading in complex networks
with degree correlations}

\author{Mari{\'a}n Bogu{\~n}{\'a}$^\star$, Romualdo Pastor-Satorras$^{\dag}$,\\
  and Alessandro  Vespignani$^{\ddag}$ \\ \ \\
  \small $^\star$Department de F{\'\i}sica Fonamental, Universitat de
  Barcelona\\ 
  \small Av. Diagonal 647, 08028 Barcelona - Spain\\
  \small $^{\dag}$Departament de F{\'\i}sica i Enginyeria Nuclear\\
  \small Universitat Polit{\`e}cnica de Catalunya, 
  \small Campus Nord,  08034 Barcelona - Spain\\
  \small $^{\ddag}$Laboratoire de Physique Th{\'e}orique (UMR du CNRS 8627)\\
  \small B{\^a}timent 210,  Universit{\'e} de Paris-Sud, 91405 ORSAY Cedex -
  France} 

\maketitle

\begin{abstract}
  We review the behavior of epidemic spreading on complex networks in
  which there are explicit correlations among the degrees of connected
  vertices.
\end{abstract}

\section{Introduction}

Complex networks arising in the modeling of many social, natural, and
technological systems are often growing and self-organizing objects
characterized by peculiar topological properties
\cite{barabasi02,dorogorev}. Many empirical evidences have prompted
that most of the times the resulting network's topology exhibits
emergent phenomena which cannot be explained by merely extrapolating
the local properties of their constituents.  Among these phenomena,
two of them appear ubiquitous in growing networks. The first one
concerns the \textit{small-world} property \cite{watts98}, that is
defined by an average path length---average distance between any pair
of vertices---increasing very slowly (usually logarithmically) with
the network size, $N$. The second one finds its manifestation in the
\textit{scale-free} (SF) degree distribution \cite{barabasi02}. This
implies that the probability $P(k)$ that a vertex has degree $k$---it
is connected to $k$ other vertices---is characterized by a power-law
behavior $P(k) \sim k^{-\gamma}$, where $2< \gamma \leq 3$ is a characteristic
exponent.

The statistical physics approach \cite{barabasi02,dorogorev} has been
proved a very valuable tool for the understanding and modeling of
these emergent phenomena in growing networks and has stimulated a more
detailed topological characterization of several social and
technological networks. In particular, it has been recognized that
many of these networks possess non-trivial degree correlations
\cite{alexei,alexei02,maslov02,assortative}. The use of statistical
physics tools has also evidentiated several surprising results
concerning dynamical processes taking place on top of complex
networks. In particular, the absence of the percolation
\cite{newman00,havlin00} and epidemic
\cite{pv01a,pv01b,lloydsir,virusreview,moreno02,newman02b} thresholds
in uncorrelated scale-free (SF) networks has hit the community because
of its potential practical implications.  The absence of the
percolative threshold, indeed, prompts to an exceptional tolerance to
random damages \cite{barabasi00}. On the other hand, the lack of any
epidemic threshold makes SF networks the ideal media for the
propagation of infections, bugs, or unsolicited information
\cite{pv01a}. While the study of uncorrelated complex networks is a
fundamental step in the understanding of the physical properties of
many systems\cite{newmanrev}, yet correlations may drastically change
the obtained results, as several recent works addressing the effect of
correlations in epidemic spreading have shown
\cite{structured,sander,blanchard,marian1,marian3}.

Here we want to provide a review of recent results concerning the
epidemic spreading in random correlated complex networks. We will
consider the sus\-cep\-ti\-ble-infected-sus\-cep\-ti\-ble (SIS) and
susceptible-infected-removed (SIR) models
\cite{anderson92,murray,epidemics} and we will provide an analytical
description that includes two vertices degree correlations in the
dynamical evolution of the infection prevalence.  This will allow us
to relate the presence or absence of epidemic threshold to the
eigenvalue spectra of certain connectivity matrices of the networks.
In particular, in the case of scale-free networks it is possible to
show that for the SIS model, a SF degree distribution $P(k)\sim k^{-\gamma}$
with $2< \gamma\leq3$ in unstructured networks with any kind of degree
correlations is a sufficient condition for a null epidemic threshold
in the thermodynamic limit.  For the SIR model, the same sufficient
condition applies if the minimum possible degree of the graph is
$k_{min}\geq2$.  The SIR model with $k_{min}=1$ has always a null
threshold unless the SF behavior is originated only by minimum degree
vertices. In other words, under very general conditions, the presence
of two-point degree correlations does not alter the extreme weakness
of SF networks to epidemic diffusion. The present results are derived
from the divergence of the nearest neighbors average degree
\cite{alexei}, which stems from the degree detailed balance condition
\cite{marian1}, to be satisfied in all physical networks.

\section{Correlated complex networks}

In the following we shall consider unstructured undirected networks,
in which all vertices within a given degree class can be considered
statistically equivalent. Thus our results will not apply to
structured networks in which a distance or time ordering can be
defined; for instance, when the small-world property is not present
\cite{klemm02,morenostructured}. We will consider in particular the
subset of undirected \textit{Markovian} random networks
\cite{marian1}, that are completely defined by the degree distribution
$P(k)$ ant the conditional probability $\condP$ that a vertex of
degree $k$ is connected to a vertex of degree $k'$. These two
functions can have any form and are assumed to be normalized, i.e.
\begin{equation}
  \sum_k P(k)= \sum_{k'} \condP = 1.
  \label{eq:5}
\end{equation}
The term ``Markovian'' refers to the fact that, in our approximation,
all higher-order correlation functions can be obtained as a
combination of the two fundamental functions $P(k)$ and $\condP$.  In
fact, this approximation represents a natural step towards a more
complex description of real networks. In this sense, the Erd{\"o}s-R{\'e}nyi
(ER) model \cite{erdos59} (defined starting from a set of $N$ vertices
that are connected in pairs with an independent probability $p$) can
be viewed as the zero-th order approximation, where the average degree
is the only fixed parameter. The ER model is thus defined by the
ensemble of all possible networks with a given average degree, but
completely random at all other respects.  The first-order
approximation has been recently introduced by realizing that many
real-world networked systems possess a more complex degree
distribution than that predicted by the ER model (a Poisson
distribution \cite{bollobas81}). In this approximation the whole
degree distribution, $P(k)$, is chosen as a constrain whereas the rest
of properties are left at random
\cite{benderoriginal,molloy95,molloy98,newmanrev}.  Even though this
approximation represents a quantitative step forward, it only takes
into account local properties and, therefore, it neglects possible
correlations among different vertices, correlations that, on the other
hand, are present in real networks \cite{alexei,assortative}. Thus, it
is quite natural to introduce the second-order approximation as that
with a fixed degree distribution, $P(k)$, and a two-point correlation
function, $\condP$, but totally random to all other respects. As we
will see, in this case the approximation must be carefully made since,
due to the two-point correlation constrain, the fundamental functions
$P(k)$ and $\condP$ must satisfy a peculiar detailed balance
condition.

The degree distribution usually identifies two kinds of networks. A
first class, which includes classical models of random graphs
\cite{erdos59}, is characterized by an exponentially bounded degree
distribution. A second one refers to SF networks in which the degree
distribution takes the form $P(k)\sim C k^{-\gamma}$, usually with $2<
\gamma\leq3$ \cite{barabasi02,dorogorev}. In this case the network shows a
very high level of degree heterogeneity, signalled by unbounded degree
fluctuations. Indeed, the second moment of the degree distribution,
$\fluck$, diverges in the thermodynamic limit $k_c\to \infty$, where $k_c$
is the maximum degree of the network. It is worth recalling that, in
growing networks, $k_c$ is related to the network size $N$ as $k_c\sim
N^{1/(\gamma-1)}$ \cite{dorogorev}. Noticeably, it is the large degree
heterogeneity of SF networks that is at the origin of their extreme
weakness towards epidemic spreading.

\subsection{Assortative and disassortative mixing}

A direct study of the conditional probability $\condP$ in data from
real networks usually yields results that are very noisy and difficult
to interpret. In order to characterize the degree correlations, it is
more useful to work with the average nearest neighbors degree (ANND)
of the vertices of degree $k$ \cite{alexei}, defined by
\begin{equation}
  \bar{k}_{nn} (k) \equiv\sum_{k'} k'\condP,
  \label{eq:6}
\end{equation}
and to plot it as a function of the degree $k$.  When two-point
correlations are not present in the network, the conditional
probability takes the form $\cprob= k' \, P(k') / \left< k \right>$,
and the ANND reads $ \bar{k}_{nn}^{nc} (k) = \left< k^2 \right> /
\left< k \right>$, which is independent on $k$. On the contrary, an
explicit dependence of $\bar{k}_{nn} (k)$ on $k$ necessary implies the
existence of non-trivial correlations, as often measured in real
networks \cite{alexei,assortative}.  For instance, in many social
networks it is observed that vertices with high degree connect more
preferably to highly connected vertices; a property referred to as
``assortative mixing''. In this case, $\bar{k}_{nn} (k)$ is an
increasing function of $k$.  On the opposite side, many technological
and biological networks show ``disassortative mixing''; i.e.  highly
connected vertices are preferably connected to vertices with low
degree and, consequently, $\bar{k}_{nn} (k)$ is a decreasing function
of $k$. Then, the ANND provides an easy and powerful way to quantify
two-point degree correlations, avoiding the fine details contained in
the full conditional probability $\condP$.

\subsection{Degree detailed balance condition}

A key relation holding for all physical networks is that all edges must point
from one vertex to another. This rather obvious observation turns out to have
important implications since it forces the fundamental functions $P(k)$ and
$\condP$ to satisfy the following degree detailed balance condition
\cite{marian1}
\begin{equation}
  k \condP P(k) = k' P(k \, \vert \, k') P(k').
  \label{detbal}
\end{equation}
This condition states that the total number of edges pointing from
vertices with degree $k$ to vertices of degree $k'$ must be equal to
the total number of edges that point from vertices with degree $k'$ to
vertices of degree $k$. This relation is extremely important since it
constraints the possible form of the conditional probability $\condP$
once $P(k)$ is given. It may be surprising that such a detailed
balance condition exists since, in fact, networks are the result of a
multiplicative random process and, in principle, detailed balance
conditions only holds for systems driven by additive noise
\cite{vankampen}. In fact, the usual detailed balance condition is the
same as Eq.~(\ref{detbal}) without the prefactors $k$ and $k'$. These
very prefactors account for the multiplicative character of the
network and Eq.~(\ref{detbal}) can be viewed as a closure condition
that guarantees the existence of the network. There is a simple way to
derive this condition. Let $N_{k}$ be the number of vertices with
degree $k$.  Obviously, $\sum_{k} N_{k}=N$ and, consequently, we can
define the degree distribution as
\begin{equation}
P(k)=\frac{N_{k}}{N}.
\end{equation}
The function $P(k)$ alone does not define completely the topology of
the network, because it says nothing about how vertices are connected
to each other. Thus, we need to define additionally the \textit{matrix
  of connections} among vertices of different degrees. Let $N_{k,k'}$
be a symmetric matrix measuring the total number of edges between
vertices of degree $k$ and vertices of degree $k'$, when $k \neq k'$,
and two times the number of self-connections, when $k=k'$. It is not
difficult to realize that this matrix fulfills the identities
\begin{eqnarray}
  \sum_{k'} N_{k,k'} &=& kN_k. \\
  \sum_{k} \sum_{k'} N_{k,k'} &=&\langle k \rangle N.
  \label{eq:1}
\end{eqnarray}
The first of this relations simply states that the number of edges
emanating from all vertices of degree $k$ is $k N_k$, while the
second indicates that the sum of all the vertices's degrees is equal
to two times the number of edges.  The identity (\ref{eq:1}) allows us
to define the joint probability
\begin{equation}
  P(k,k')=\frac{N_{k,k'}}{\langle k \rangle N},
\end{equation}
where the symmetric function $(2-\delta_{k,k'})P(k,k')$ is the probability
that a randomly chosen edge connects two vertices of degrees $k$ and
$k'$. The correlation coefficient computed from this joint probability
has been recently used in Ref.~\cite{assortative} in order to quantify
two-point degree correlations. The transition probability $\condP$,
defined as the probability that an edge from a $k$ vertex points to a
$k'$ vertex, can be easily written as
\begin{equation}
P(k' \, \vert \, k)= \frac{N_{k' , k}}{k N_k} \equiv 
\frac{\avk P(k,k')}{k P(k)},
\end{equation}
from where the detailed balance condition arises as a consequence of
the symmetry of $P(k,k')$ (or $N_{k,k'}$).

From the degree detailed balance condition it is possible to derive
some general exact results concerning the behavior of $\bar{k}_{nn}
(k, k_c)$ and of $\left<\bar{k}_{nn}\right>_{k_c}=\sum_kP(k)
\bar{k}_{nn} (k, k_c)$ in SF networks \cite{marian3}. In these two
functions we have now made explicit the $k_c$ dependence originated by
the upper cut-off of the $k$ sum and that must be taken into account
since it is a possible source of divergences in the thermodynamic
limit.  The results that we will derive will turn out to be
fundamental in determining the epidemic spreading behavior in these
networks.  Let us start by multiplying by a $k$ factor both terms of
Eq.~(\ref{detbal}) and summing over $k'$ and $k$.  We obtain
\begin{equation}
 \fluck  = \sum_{k'}k'P(k')\sum_k kP(k \, \vert \, k'),
\label{detbal2}
\end{equation}
In SF networks with $2< \gamma < 3$ we have that the second moment of the
degree distribution diverges as $\fluck\sim k_c^{3-\gamma}$ \footnote{For
  $\gamma=3$ the second moment diverges as $\fluck \sim \ln k_c$ but the
  argument, though more involved, is still valid.}.  We thus obtain
from Eq.~(\ref{detbal2}), using the definition (\ref{eq:6}),
\begin{equation}
  \sum_{k'}k'P(k') \bar{k}_{nn}(k',k_c)
  \simeq \frac{C}{(3-\gamma)} k_c^{3-\gamma},
\label{detbal3}
\end{equation}
where $C$ is the constant prefactor from the degree distribution.  In
the case of disassortative mixing \cite{assortative}, the function
$\bar{k}_{nn}(k',k_c)$ is decreasing with $k'$ and, since $k'P(k')$ is
an integrable function, the l.h.s. of Eq.~(\ref{detbal3}) has no
divergence related to the sum over $k'$. This implies that the
divergence must be contained in the $k_c$ dependence of
$\bar{k}_{nn}(k',k_c)$.  In other words, the function
$\bar{k}_{nn}(k',k_c)\to\infty$ for $k_c\to\infty$ in a non-zero measure set. In
the case of assortative mixing, $\bar{k}_{nn}(k',k_c)$ is an
increasing function of $k'$ and, depending on its rate of growth,
there may be singularities associated to the sum over $k'$.
Therefore, this case has to be analyzed in detail. Let us assume that
the ANND grows as $\bar{k}_{nn}(k',k_c) \simeq \alpha k'^{\beta}$, $\beta>0$, when
$k' \to \infty$. If $\beta <\gamma -2$, again there is no singularity related to
the sum over $k'$ and the previous argument for disassortative mixing
holds.  When $\gamma -2 \leq \beta <1$ there is a singularity coming from the
sum over $k'$ of the type $\alpha k_c^{\beta-(\gamma-2)}$. However, since
Eq.~(\ref{detbal3}) comes from an identity, the singularity on the
l.h.s. must match both the exponent of $k_c$ and the prefactor on the
r.h.s.  In the case $\gamma -2 \leq \beta <1$, the singularity coming from the
sum is not strong enough to match the r.h.s. of Eq.~(\ref{detbal3})
since $\beta-(\gamma-2) < 3-\gamma$. Thus, the function $\bar{k}_{nn}(k',k_c)$
must also diverge when $k_c \to \infty$ in a non-zero measure set. Finally,
when $\beta >1$ the singularity associated to the sum is too strong,
forcing the prefactor to scale as $\alpha \simeq r k_c^{1-\beta}$ and the ANND as
$\bar{k}_{nn}(k',k_c)\simeq r k_c^{1-\beta} k'^{\beta}$. It is easy to realize
that $r \leq 1$, since the ANND cannot be larger than $k_c$. Plugging
the $\bar{k}_{nn}(k',k_c)$ dependence into Eq.~(\ref{detbal3}) and
simplifying common factors, we obtain the identity at the level of
prefactors
\begin{equation}
  \frac{r}{2-\gamma+\beta} = \frac{1}{3-\gamma}.
  \label{eq:pref}
\end{equation}
Since $\beta>1$ and $r<1$, the prefactor in the l.h.s. of
Eq.~(\ref{eq:pref}) is smaller than the one of the r.h.s. This fact
implies that the tail of the distribution in the l.h.s. of
Eq.~(\ref{detbal3}) cannot account for the whole divergence of its
r.h.s. This means that the sum is not the only source of divergences
and, therefore, the ANND must diverge at some other point.

In summary, the function $\bar{k}_{nn}(k',k_c)$ must diverge when $k_c
\to \infty$ in a non-zero measure set independently of the correlation
behavior.  The large $k_c$ singularity of the ANND can then be used to
evaluate the quantity
\begin{equation}
  \left< \bar{k}_{nn}\right>_{N}=\sum_k P(k) \bar{k}_{nn}(k ,k_c),
\end{equation}
where we have explicitly considered $k_c$ as a growing function of the
network size $N$. The r.h.s. of this equation is a sum of positive
terms and diverges with $k_c$ at least as $\bar{k}_{nn}(k ,k_c)$ both
in the disassortative or assortative cases. In other words, {\em all
  SF networks with $2<\gamma\leq3$ must present a $\left<
    \bar{k}_{nn}\right>_{N}\to\infty$ for $N\to\infty$}.  This statement is
independent of the structure of the correlations present in the
network.  The quantity $\left< \bar{k}_{nn}\right>_{N}$ is
particularly useful in model analysis and real data measurements.
Degree correlation functions can be measured in several networks, but
measurements are always performed in the presence of a finite $k_c$
that does not allow to exploit the singularity of the function
$\bar{k}_{nn}(k,k_c)$.  The most convenient way to exploit the
infinite size singularity is therefore to measure the $\left<
  \bar{k}_{nn}\right>_{N}$ for increasing network sizes.

The quantity $\left< \bar{k}_{nn}\right>_{N}$ is very important in
defining the properties of spreading processes in networks since it
measures the number of individuals that can be infected in a few
contagions.  We shall discuss this point in relation to some specific
epidemic models in the next sections.

\section{The SIS model}

As a first prototypical example for examining the properties of
epidemic dynamics in SF networks we consider the
susceptible-infected-susceptible (SIS) model \cite{anderson92}, in
which each vertex represents an individual of the population and the
edges represent the physical interactions among which the infection
propagates.  Each individual can be either in a susceptible or
infected state.  Susceptible individuals become infected with
probability $\lambda$ if at least one of the neighbors is infected.
Infected vertices, on the other hand, recover and become susceptible
again with probability one. A different recovery probability can be
considered by a proper rescaling of $\lambda$ and the time. This model is
conceived for representing endemic infections which do not confer
permanent immunity, allowing individuals to go through the stochastic
cycle susceptible $\to$ infected $\to$ susceptible by contracting the
infection over and over again.

\subsection{Uncorrelated homogeneous networks}

In uncorrelated homogeneous networks, in which each vertex has more or
less the same number of connections, $k\simeq\avk$, a general result
states the existence of a finite epidemic threshold, separating an
infected (endemic) phase, with a finite average density of infected
individuals, from a healthy phase, in which the infection dies out
exponentially fast \cite{epidemics}. This is for instance the case of
random networks with exponentially bounded degree distribution.

This result can be recovered by considering the dynamical evolution of
the average density of infected individuals $\rho(t)$ (the prevalence)
present in the network. The SIS model in homogeneous uncorrelated
networks at a mean-field level is described by the following rate
equation \cite{pv01b}
\begin{equation}
  \frac{d \rho(t)}{d t} = -\rho(t) +\lambda \avk \rho(t) \left[ 1-\rho(t) \right].
\label{eq:ws}
\end{equation}
In this equation we have neglected higher order terms, since we are
interested in the onset of the endemic state, close to the point
$\rho(t) \sim 0$. Also, we have neglected correlations among vertices.
That is, the probability of infection of a new vertex---the second
term in Eq.~(\ref{eq:ws})---is proportional to the infection rate
$\lambda$, to the probability that a vertex is healthy, $1-\rho(t)$, and to
the probability that a edge in a healthy vertex points to an infected
vertex.  This last quantity, assuming the \textit{homogeneous mixing
  hypothesis}\footnote{The homogeneous mixing hypothesis
  \cite{anderson92} states that the force of the infection (the
  \textit{per capita} rate of acquisition of the disease by the
  susceptible individuals) is proportional to the density of infected
  individuals.}, is approximated for homogeneous networks as $\avk
\rho(t)$, i.e. proportional to the average number of connections and to
the density of infected individuals.  From Eq.~(\ref{eq:ws}) it can be
proved the existence of an epidemic threshold $\lambda_c= \avk^{-1}$
\cite{epidemics}, such that $\rho = 0$ if $\lambda< \lambda_c$, while $\rho \sim
(\lambda-\lambda_c)$ if $\lambda\geq \lambda_c$.  In this context, it is easy to recognize
that the SIS model is a generalization of the \textit{contact process}
model, widely studied as the paradigmatic example of an
absorbing-state phase transition to a unique absorbing state
\cite{marro99}.

\subsection{Uncorrelated complex networks}

For general complex networks, in which large degree fluctuations
and correlations might be allowed, we must relax the homogeneous
hypothesis made in writing Eq.~(\ref{eq:ws}) and work instead with the
relative density $\rho_k(t)$ of infected vertices with given
degree $k$; i.e.  the probability that a vertex with $k$ edges
is infected.  Following Refs.~\cite{pv01a,pv01b}, the rate equation
for $\rho_k(t)$ can be written as
\begin{equation}
  \frac{ d \rho_k(t)}{d t} =
  -\rho_k(t) +\lambda k \left[1-\rho_k(t) \right]  \Theta_k(t).
\label{mfk}
\end{equation}
In this case, the creation term is proportional to the spreading rate
$\lambda$, the density of healthy sites $1-\rho_k(t)$, the degree $k$,
and the variable $\Theta_k(t)$, that stands for the probability that an
edge emanating from a vertex of degree $k$ points to an infected
site.  In the case of an uncorrelated random network, considered in
Refs.~\cite{pv01a,pv01b}, the probability that a edge points to a
vertex with $k$ connections is equal to $kP(k)/ \avk$
\cite{newmanrev}. This yields a $\Theta_k=\Theta^{\rm nc}$ independent of $k$
that reads as
\begin{equation}
  \Theta^{\rm nc} =\frac{1}{\avk}\sum_{k'} k' P(k')\rho_{k'}(t).
  \label{first}
\end{equation}
Substituting the expression (\ref{first}) into Eq.~(\ref{mfk}), it is
possible to find the steady state solution where $\Theta^{\rm nc}$ is now
a function of $\lambda$ alone~\cite{pv01a,pv01b} by the following
self-consistent equation:
\begin{equation}
  \Theta^{\rm nc} = \frac{1}{\avk}  \sum_k k P(k)  \frac{\lambda k
\Theta^{\rm nc}}{1 + \lambda k \Theta^{\rm nc}}.
\label{cons}
\end{equation}
A non-zero stationary prevalence ($\rho_k\neq 0$) is obtained when the
r.h.s. and the l.h.s.  of Eq.~(\ref{cons}), expressed as a function of
$\Theta^{\rm nc}$, cross in the interval $0<\Theta^{\rm nc}\leq 1$, allowing a
nontrivial solution.  It is easy to realize that this corresponds to
the inequality
\begin{equation}
  \frac{d}{d \Theta^{\rm nc} } \left. \left( \frac{1}{\avk}
      \sum_k k P(k)  \frac{\lambda k
        \Theta^{\rm nc}}{1 + \lambda k \Theta^{\rm nc}}
    \right) \right|_{\Theta^{\rm nc} =0} \geq 1
  \label{eq:critpunt}
\end{equation}
being satisfied. The value of $\lambda$ yielding the equality in
Eq.~(\ref{eq:critpunt}) defines the critical epidemic threshold
$\lambda_c$, that is given for uncorrelated random networks by
\begin{equation}
  \lambda_c^{\rm nc} =\frac{\avk}{\fluck}.
  \label{eq:2}
\end{equation}
In uncorrelated and infinite SF networks with $\gamma \leq 3$, we therefore
have $\fluck = \infty$, and correspondingly $\lambda_c^{\rm nc} = 0$. This is a
very relevant result, signalling that the high heterogeneity of SF
networks makes them extremely weak with respect to infections.  These
results have several implications in human and computer virus
epidemiology \cite{lloyd01}.

\subsection{Correlated complex networks}

For a general network in which the degrees of the vertices are
correlated, the above formalism is not correct, since we are not
considering the effect of the degree $k$ into the expression for
$\Theta_k$. This effect can be taken into account, however, for Markovian
networks, whose correlations are completely defined by the conditional
probability $\condP$.  In this case, it is easy to realize that the
correct factor $\Theta_k$ can be written as
\begin{equation}
  \Theta_k(t) = \sum_{k'} \condP \rho_{k'}(t),
  \label{eq:3}
\end{equation}
that is, the probability that an edge in a vertex of degree $k$ is
pointing to an infected vertex is proportional to the probability that
any edge points to a vertex with degree $k'$, times the probability
that this vertex is infected, $\rho_{k'}(t)$, averaged over all the
vertices connected to the original vertex. Eqs.~(\ref{mfk})
and~(\ref{eq:3}) define together the mean-field equation describing the SIS
model on Markovian complex networks,
\begin{equation}
  \frac{ d \rho_k(t)}{d t} =
  -\rho_k(t) +\lambda k \left[1-\rho_k(t) \right]
\sum_{k'} \condP \rho_{k'}(t).
  \label{generalized}
\end{equation}
It must be stressed that this equation is valid only if the network
has no structure; i.e. the only relevant variable is the degree $k$.
This implies that all vertices within a given degree class are
statistically equivalent. This is not the case, for instance, of
regular lattices or structured networks
\cite{klemm02,morenostructured} in which a spatial ordering is
constraining the connectivity among vertices.  The exact solution of
Eq.~(\ref{generalized}) can be difficult to find, depending on the
particular form of $\condP$. However, it is possible to extract the
value of the epidemic threshold by analyzing the stability of the
steady-state solutions. Of course, the healthy state $\rho_k=0$ is one
solution. For small $\rho_k$, we can linearize Eq.~(\ref{generalized}),
getting
\begin{equation}
  \frac{ d \rho_k(t)}{d t} \simeq \sum_{k'} L_{k  k'} \rho_{k'}(t).
\end{equation}
In the previous equation we have defined the Jacobian matrix
$\mathbf{L}=\{ L_{k  k'} \}$ by
\begin{equation}
  L_{k  k'} = -\delta_{k  k'} + \lambda k \condP,
\end{equation}
where $\delta_{k k'}$ is the Kronecker delta symbol. The solution
$\rho_k=0$ will be unstable if there exists at least one positive
eigenvalue of the Jacobian matrix $\mathbf{L}$.  Let us consider the
\textit{connectivity matrix} $\mathbf{C}$, defined by $C_{k k'} = k
\condP$. Using the symmetry condition Eq.~(\ref{detbal}), it is easy to
check that if $v_k$ is an eigenvector of $\mathbf{C}$, with eigenvalue
$\Lambda$, then $P(k) v_k$ is an eigenvector of the transposed matrix
$\mathbf{C}^T$ with the same eigenvalue. From here it follows
immediately that all the eigenvalues of $\mathbf{C}$ are real.
Let $\Lambda_m$ be the
largest eigenvalue of  $\mathbf{C}$.
Then, the origin will be unstable whenever $- 1 + \lambda \Lambda_m >0$, which
defines the epidemic threshold
\begin{equation}
  \lambda_c = \frac{1}{ \Lambda_m},
  \label{eq:4}
\end{equation}
above which the solution $\rho_k=0$ is unstable, and another non-zero
solution takes over as the actual steady-state---the endemic state.

It is instructive to see how this general formalism recovers previous
results \cite{pv01a,pv01b}, implicitly obtained for random
uncorrelated networks. For any random network, in which there are no
correlations among the degrees of the vertices, we have that the
connectivity matrix is given by $C_{k k'}^{\rm nc} = k P(k' / k) \equiv k
k' P(k')/ \avk$, since the probability that an edge points to a vertex
of connectivity $k'$ is proportional to $k' P(k')$. It is easy to
check that the matrix $\{ C_{k' k}^{\rm nc} \} $ has a unique
eigenvalue $\Lambda_m^{\rm nc} = \fluck / \avk$, corresponding to the
eigenvector $v_k^{\rm nc} = k$, from where we recover the now
established result Eq.~(\ref{eq:2}).

\subsection{Correlated Scale-Free networks}

The absence of an epidemic threshold in SF uncorrelated networks is an
extremely important question that prompts to a possible weakness of
many real-world networks. Particularly important is for the case of
digital viruses spreading on the Internet \cite{pv01a,alexei} and
sexually transmitted diseases diffusing on the web of sexual contacts
\cite{amaral01}.  Both these networks show, in fact, SF properties
that would imply the possibility of major epidemic outbreaks even for
infections with a very low transmission rate. Immunization policies
as well must be radically changed in the case that a network has a
null epidemic threshold \cite{psvpro,aidsbar,havlinimmuno}.

In view of the relevance of this framework, it is extremely important
to study to which extent the presence of correlations are altering
these results. The main question is therefore which conditions on the
degree correlations of SF networks preserve the lack of a critical
threshold. In the case of correlated networks, we have shown that the
epidemic threshold is the inverse of the largest eigenvalue of the
connectivity matrix $\mathbf{C}$. The absence of an epidemic threshold
thus corresponds to a divergence of the largest eigenvalue of the
connectivity matrix $\mathbf{C}$ in the limit of an infinite network
size $N\to\infty$.  In order to provide some general statement on the
conditions for such a divergence we can make use of the Frobenius
theorem for non-negative irreducible matrices \cite{Gantmacher}. This
theorem states the existence of the largest eigenvalue of any
non-negative irreducible matrix, eigenvalue which is simple, real,
positive, and has a positive eigenvector. In our case the matrix of
interest is the connectivity matrix that is non-negative and
irreducible.  The irreducible property of the connectivity matrix is a
simple consequence of the fact that all the degree classes in
the network are accessible. That is, starting from the degree
class $k$ it is always possible to find a path of edges that connects
this class to any other class $k'$ of the network. If this is not the
case it means that the network is built up of disconnected irreducible
subnetworks and, therefore, we can apply the same line of reasoning to
each subnetwork\footnote{Notice that being irreducible is not
  equivalent to being fully connected at the vertex to vertex level,
  but at the class to class level.}.  From the Frobenius theorem
\cite{Gantmacher} it can be proved that the maximum eigenvalue,
$\Lambda_m$, of any non-negative irreducible matrix, $A_{k k'}$, satisfies
the inequality
\begin{equation}
  \Lambda_m \geq \min_k \frac{1}{\psi(k)}\sum_{k'} A_{k k'}
\psi(k'),
\label{eq:10}
\end{equation}
where $\{ \psi(k) \}$ is any positive vector. In particular, by setting
${\mathbf A}={\mathbf C^2}$ and $\psi(k)=k$ we obtain the inequality
\begin{equation}
\Lambda_m^2\geq \min_k \sum_{k'}\sum_{\ell}k'\ell
\condPN{\ell }{k} \condPN{k'}{\ell}.
\label{frob}
\end{equation}
This inequality relates the lower bound of the largest eigenvalue
$\Lambda_m$ to the degree correlation function and
allows to find a sufficient condition for the absence of the epidemic
threshold. By noting that
$\sum_{k'}k'\condPN{k'}{\ell}=\bar{k}_{nn}(\ell ,k_c)$,
we obtain the inequality
\begin{equation}
\Lambda_m^2\geq \min_k\sum_{\ell} \ell \condPN{\ell}{k}
\bar{k}_{nn}(\ell ,k_c).
\label{frob2}
\end{equation}
The r.h.s. of this equation is a sum of positive terms, and by
recalling the divergence of the ANND with $k_c$, we readily obtain
that $\Lambda_m\geq\infty$ for all networks with diverging $\fluck$ both in the
disassortative or assortative cases\footnote{One may argue that, since
  we are calculating a minimum for $k$, if the transition probability
  $P(\ell \, | \, k_0)$ is zero at some point $k_0$, this minimum is
  zero. In this case it is possible to show that repeating the same
  argument with ${\mathbf C}^3$ instead of ${\mathbf C}^2$ provides us
  an inequality that avoids this problem, Ref.~\cite{marian1}.}.  The
divergence of $\Lambda_m$ implies on its turn that the SIS epidemic
threshold vanishes, in the thermodynamic limit, in all SF networks
with assortative or disassortative mixing if the degree distribution
has a diverging second moment; {\em i.e. a SF degree distribution with
  exponent $2<\gamma\leq3$ is a sufficient condition for the absence of an
  epidemic threshold for the SIS model in unstructured networks with
  arbitrary two-point degree correlation function}.
 
In physical terms, the absence of the epidemic threshold is related to
the divergence of  $\left<\bar{k}_{nn}\right>_{N}$ in SF networks.
In homogeneous networks, where $\left< \bar{k}_{nn}\right>_{N}\simeq\avk$,
the epidemic spreading properties can be related to the average
degree. In SF networks, however, the focus shifts to the possibility
of infecting a large number of individuals in a finite number of
contagions. The fact that an infected vertex has a very low degree is
not very important if a hub of the network that provides connectivity
to a large number of vertices is a few hops away.  The infection can,
in this case, very easily access a very large number of individuals in
a short time.  In SF networks is the ANND that takes into account more
properly the level of degree fluctuations and thus rules the epidemic
spreading dynamics.

It is worth stressing that the divergence of $\left<
  \bar{k}_{nn}\right>_{N}$ is ensured by the degree detailed balance
condition alone, and it is a very general result holding for all SF
networks with $2<\gamma\leq3$.  On the contrary, the SF behavior with
$2<\gamma\leq3$ is a sufficient condition for the lack of epidemic threshold
only in networks with general two-point degree correlations and in
absence of higher-order correlations.  The reason is that the relation
between the epidemic threshold and the maximum eigenvalue of the
connectivity matrix only holds for these classes of networks. Higher
order correlations, or the presence of an underlying metric in the
network \cite{morenostructured}, can modify the rate equation at the
basis of the SIS model and may invalidate the present discussion.

\section{The SIR model}

The susceptible-infected-removed model (SIR) \cite{anderson92}
represents the other paradigmatic example of epidemic dynamics. Unlike
in the SIS model, in this case infected individuals fall, after some
random time, into a removed state where they cannot neither become
infected again nor infect other individuals. This model tries to
mimic real epidemics where individuals, after being infected, acquire
permanent immunity or, in the worst case, die.

The SIR model is defined as follows. Individuals can only exist in
three different states, namely, susceptible, infected, or removed.
Susceptible individuals become infected with probability $\lambda$ if at
least one of their neighbors is infected. On the other hand, infected
individuals spontaneously fall in the removed state with probability
$\mu$, which without lack of generality we set equal to unity.  The
main difference between both models is that whereas in the SIS, for
$\lambda>\lambda_c$, the epidemics reaches a steady state, in the SIR the
epidemics always dies and reaches eventually a state with zero density
of infected individuals. The epidemic prevalence is thus defined in
this case as the total number of infected individuals in the whole
epidemic process.

\subsection{Uncorrelated homogeneous networks}

In a homogeneous system, the SIR model can be described in terms of
the densities of susceptible, infected, and removed individuals,
$S(t)$, $\rho(t)$, and $R(t)$, respectively, as a function of time.
These three quantities are linked through the normalization condition
\begin{equation}
  S(t) + \rho(t) + R(t) =1,
  \label{eq:norm}
\end{equation}
and they obey the
following system of differential equations:
\begin{eqnarray}
  \frac{d S}{d t} &=& - \lambda \avk \rho S, \nonumber \\
  \frac{d \rho}{d t} &=& - \rho + \lambda \avk 
   \rho S,  \label{eq:primiser}\\
  \frac{d R}{d t} &=&  \rho. \nonumber 
\end{eqnarray}
These equations can be interpreted as follows: infected individuals
decay into the removed class at a unity rate, while susceptible
individuals become infected at a rate proportional to both the
densities of infected and susceptible individuals.  Here, $\lambda$ is the
microscopic spreading (infection) rate, and $\avk$ is the number of
contacts per unit time that is supposed to be constant for the whole
population. In writing this last term of the equations we have assumed
again, as in the case of the SIS model, the homogeneous mixing
hypothesis \cite{anderson92},

The most significant prediction of this model is the presence of a
nonzero epidemic threshold $\lambda_c$ \cite{murray}.  If the value of $\lambda$
is above $\lambda_c$, $\lambda>\lambda_c$, the disease spreads and infects a finite
fraction of the population. On the other hand, when $\lambda$ is below the
threshold, $\lambda<\lambda_c$, the total number of infected individuals (the
epidemic prevalence), $\Ri = \lim_{t\to\infty} R(t)$, is infinitesimally
small in the limit of very large populations. In order to see this
point, let us consider the set of equations (\ref{eq:primiser}).
Integrating the equation for $S(t)$ with the initial conditions
$R(0)=0$ and $S(0)\simeq1$ (i.e., assuming $\rho(0) \simeq 0$, a very small
initial concentration of infected individuals), we obtain
\begin{equation}
S(t) = e^{-\lambda \avk R(t)}.  
\end{equation}
Combining this result with the normalization condition
(\ref{eq:norm}), we observe that the total number of infected
individuals $\Ri$ fulfills the following self-consistent equation:
\begin{equation}
  \Ri = 1 - e^{-\lambda \avk \Ri}.  
\end{equation}
While $\Ri=0$ is always a solution of this equation, in order to have
a nonzero solution the following condition must be fulfilled:
\begin{equation}
    \frac{d}{d \Ri } \left. \left( 1  -   
e^{ -\lambda \avk \Ri} \right) \right|_{\Ri=0} \geq  1. 
\end{equation}
This condition is equivalent to the constraint $\lambda \geq \lambda_c$, where the
epidemic threshold $\lambda_c$ takes the value $\lambda_c=\avk^{-1}$.
Performing a Taylor expansion at $\lambda=\lambda_c$ it is then possible to
obtain the epidemic prevalence behavior $\Ri\sim(\lambda -\lambda_c)$ (valid above
the epidemic threshold). From the point of view of the physics of
non-equilibrium phase transition, it is easy to recognize that the SIR
model is a generalization of the \textit{dynamical percolation} model,
that has been extensively studied in the context of absorbing-state
phase transitions \cite{marro99}.

\subsection{Uncorrelated complex networks}

In order to take into account the heterogeneity induced by the
presence of vertices with different degree, we consider the time
evolution of the magnitudes $\rho_k(t)$, $S_k(t)$, and $R_k(t)$, which
are the density of infected, susceptible, and removed vertices of
degree $k$ at time $t$, respectively \cite{moreno02,newman02b}. These variables
are connected by means of the normalization condition
\begin{equation}
  \rho_k(t)+S_k(t)+R_k(t) = 1.
\end{equation}
Global quantities such as the epidemic prevalence can be expressed as
an average over the various degree classes; for example, we define the
total number of removed individuals at time $t$ by $R(t)=\sum_k
P(k)R_k(t)$, and the prevalence as $\Ri = \lim_{t \to \infty} R(t)$.  At
the mean-field level, for random uncorrelated networks, these
densities satisfy the following set of coupled differential equations:
\begin{eqnarray}
  \frac{d \rho_k(t)}{d t} & = & -\rho_k(t) +\lambda k S_k(t) \Theta^{\rm nc}(t), \label{eq:sir1}\\
  \frac{d S_k(t)}{d t} & = & - \lambda k S_k(t) \Theta^{\rm nc}(t), \label{eq:sir2}\\
  \frac{d R_k(t)}{d t} & = & \rho_k(t). \label{eq:sir3}
\end{eqnarray}
The factor $\Theta^{\rm nc}(t)$ represents the probability that any given edge
points to an infected vertex and is capable of transmitting the
disease. This quantity can be computed in a self-consistent way: The
probability that an edge points to an infected vertex with degree $k'$
is proportional to $k' P(k')$. However, since the infected vertex
under consideration received the disease through a particular edge
that cannot be used for transmission anymore (since it points back to
a previously infected individual) the correct probability must
consider one less edge. Therefore,
\begin{equation}
  \Theta^{\rm nc}(t) = \frac{1}{\avk} \sum_k (k-1) P(k) \rho_k(t).
  \label{eq:definition}
\end{equation}
The equations (\ref{eq:sir1}), (\ref{eq:sir2}), (\ref{eq:sir3}), and
(\ref{eq:definition}), combined with the initial conditions
$R_k(0)=0$, $\rho_k(0) = \rho_k^0$, and $S_k(0) = 1- \rho_k^0$, completely
define the SIR model on any uncorrelated complex network with degree
distribution $P(k)$.  We will consider in particular the case of a
homogeneous initial distribution of infected individuals, $\rho_k^0 =
\rho^0$.  In this case, in the limit $\rho^0 \to 0$, we can substitute
$\rho_k(0) \simeq 0$ and $S_k(0) \simeq 1$. Under this approximation,
Eqs.~\equ{eq:sir2} and (\ref{eq:sir3}) can be directly integrated,
yielding
\begin{equation}
  S_k(t) =  e^{ -\lambda k \phi(t)}, \qquad R_k(t) = \int_0^\infty
  \rho_k(\tau) d t,
  \label{eq:sir4}
\end{equation}
where we have defined the auxiliary function 
\begin{equation}
  \phi(t) = \int_0^t \Theta^{\rm nc}(\tau) d \tau = \frac{1}{\avk} \sum_k
  (k-1) P(k) R_k(t). 
  \label{eq:sir5}
\end{equation}
 
In order to get a closed relation for the total density of infected
individuals, it results more convenient to focus on the time evolution
of the averaged magnitude $\phi(t)$. To this purpose, let us compute its
time derivative:
\begin{equation}
  \frac{ d \phi(t)}{d t} 
  = 1 - \frac{1}{\avk} - \phi(t) -  \frac{1}{\avk} \sum_k (k-1) P(k) e^{
    -\lambda k \phi(t)} \label{eq:generalphi}, 
\end{equation}
where we have introduced the time dependence of $S_k(t)$ obtained in
Eq.~(\ref{eq:sir4}). Once solved Eq.~(\ref{eq:generalphi}), we can
obtain the total epidemic prevalence $\Ri$ as a function of $\phi_\infty =
\lim_{t \to \infty} \phi(t)$. Since $R_k(\infty) = 1 - S_k(\infty)$, we have
\begin{equation}
  \Ri = \sum_k P(k) \left(1 - e^{ -\lambda k \phi_\infty }\right).
  \label{eq:rigeneral}
\end{equation}

For a general $P(k)$ distribution, Eq.~\equ{eq:generalphi} cannot be
generally solved in a closed form. However, we can still get useful
information on the infinite time limit; i.e. at the end of the
epidemics. Since we have that $\rho_k(\infty)=0$, and consequently $\lim_{t
  \to \infty} \mathrm{d} \phi(t) / \mathrm{d} t = 0$, we obtain from
Eq.~\equ{eq:generalphi} the following self-consistent equation for
$\Fi$:
\begin{equation}
    \phi_\infty  = 1  - \frac{1}{\avk}- \frac{1}{\avk} \sum_k (k-1) P(k) e^{
      -\lambda k \phi_\infty}. 
    \label{eq:sirselfcons}
\end{equation}
The value $ \phi_\infty = 0$ is always a solution. In order to have a non-zero
solution, the condition 
\begin{equation}
    \frac{d }{d \phi_\infty } \left. \left( 1  -  \frac{1}{\avk} -
        \frac{1}{\avk} \sum_k (k-1) P(k) e^{ 
          -\lambda  k \phi_\infty} \right) \right|_{\phi_\infty=0} \geq  1
\end{equation}
must be fulfilled. This relation implies
\begin{equation}
 \frac{\lambda}{\avk} \sum_k k (k-1) P(k) \geq  1,
\end{equation}
which defines the epidemic threshold
\begin{equation}
  \lambda^{\rm nc}_c =  \frac{\avk}{\fluck -\avk},
  \label{eq:sirthreshold}
\end{equation}
below which the epidemic prevalence is null, and above which it
attains a finite value. It is interesting to notice that this is
precisely the same value found for the percolation threshold in
generalized networks \cite{newman00,havlin00}. This is hardly
surprising since, as it is well known \cite{grassbergerperc}, the SIR
model can be mapped to a bond percolation process.

\subsection{Correlated complex networks}

In order to work out the SIR model in Markovian networks it is easier
to consider the rate equations for the quantities $N_k^I (t)$ and
$N_k^R (t)$, defined as the number of infected and removed individuals
of degree $k$, present at time $t$, respectively. From this two
quantities we can easily recover the densities $\rho_k(t)$ and $R_k(t)$
as
\begin{equation}
  \rho_k(t) = \frac{N_k^I (t)}{N_k}, \qquad R_k(t) = \frac{N_k^R
    (t)}{N_k}, \qquad S_k(t) = 1 -  \rho_k(t) - R_k(t),
\end{equation}
where $N_k = NP(k)$ is the number of vertices with degree $k$. The
rate equations for $N_k^I (t)$ and $N_k^R (t)$ are then given by
\begin{eqnarray}
  \frac{d N_k^I (t)}{dt} & = & -N_k^I(t)+\lambda
 S_k(t)  \Gamma_k(t), \label{sirN} \\
 \frac{d N_k^R (t)}{dt} & = & N_k^I(t),
\end{eqnarray}
where we have defined the function
\begin{equation}
\Gamma_k(t)\equiv \sum_{k'} N_{k'}^I(t) (k'-1) P(k \, \vert \, k')
\end{equation}
In this case, the creation of new infected individuals---the second
term in the r.h.s. of Eq.~(\ref{sirN})---is proportional to the number
of infected individuals of degree $k'$, $N_{k'}^I(t)$, the probability
that a vertex of degree $k$ is susceptible, $S_k(t)$, and the average
number of edges pointing from these infected vertices to vertices of
degree $k$, $(k'-1) P(k \, \vert \, k')$, all averaged for all the
vertices of degree $k'$. This last term takes into account that one of
the edges is not available for transmitting the infection, since it
was used to infect the vertex considered.  Dividing Eq.~(\ref{sirN})
by $N_k$ and making use of the detailed balance condition
Eq.~(\ref{detbal}) we find the rate equations for the relative
densities as
\begin{eqnarray}
\frac{d R_k(t) (t)}{dt} &=& \rho_k(t), \\
\frac{d \rho_k (t)}{dt} &=& -\rho_k(t)+\lambda k S_k(t)  \Theta_k(t),
\label{eq:7}
\end{eqnarray}
where, for a Markovian network, the factor $\Theta_k(t)$ takes the form
\begin{equation}
\Theta_k(t)=\sum_{k'} \rho_{k'}(t) \frac{k'-1}{k'} P(k' \, \vert \, k)
\label{eq:8}
\end{equation}
Again, it must be stressed that no structure is allowed in the network
in order for this equations to represent a valid formulation of the
SIR model.

In order to extract information about the epidemic threshold, we
proceed similarly to the SIS model, performing a linear stability
analysis. For time $t \to 0$, that corresponds to small $\rho_k$, $R_k \simeq
0$ and $S_k \simeq 1$, Eqs.~(\ref{eq:7}) and~(\ref{eq:8}) can be written
as
\begin{equation}
  \frac{d \rho_k (t)}{dt} \simeq \sum_{k'} \tilde{L}_{k k'} \rho_{k'}(t),
  \label{eq:9}
\end{equation}
where the Jacobian matrix $\mathbf{\tilde{L}} = \{ \tilde{L}_{k k'}
\}$ can be written as
\begin{equation}
  \tilde{L}_{k k'} = -\delta_{k k'} + \lambda \frac{k(k'-1)}{k'}  P(k' \, \vert \, k).
\end{equation}
In order to infect a finite fraction of individuals, we need the
solution $\rho_k=0$ to be unstable, which happens if there is at least
one positive eigenvalue of the Jacobian matrix $\mathbf{\tilde{L}}$.
Defining the matrix $\mathbf{\tilde{C}} = \{ \tilde{C}_{k k'} \}$,
with elements
\begin{equation}
  \tilde{C}_{k k'}  = \frac{k(k'-1)}{k'}  P(k' \, \vert \, k),
\end{equation}
we know from the Frobenius theorem, since it is positive and provided
that it is reducible at the degree class level, that it has a largest
eigenvalue $\tilde{\Lambda}_m$ that is real and positive. Thus, the
solution $\rho_k=0$ of Eq.~(\ref{eq:9}) is stable whenever $-1+ \lambda
\tilde{\Lambda}_m  < 0$. This relation defines the epidemic threshold for the
SIR model in Markovian networks
\begin{equation}
  \lambda=\frac{1}{\tilde{\Lambda}_m}.
\end{equation}
In the case of a random uncorrelated network, we have that
$\tilde{C}^{\rm nc}_{k k'} = k (k'-1) P(k')/ \avk$. It can be easily
seen that this matrix has a unique eigenvalue $\tilde{\Lambda}_m^{\rm nc} =
\fluck/ \avk -1$, corresponding to the eigenvalue $\tilde{v}^{\rm nc}_k =k$,
thus recovering the previous result Eq.~(\ref{eq:sirthreshold})
obtained for this kind of networks.

The relation between the SIR model and percolation in correlated
complex networks can be closed by noticing that the relevant parameter
in this last problem is the largest eigenvalue of the matrix
$\mathbf{C}^{\rm perc} = \{ {C}^{\rm perc}_{k k'} \}$, with elements
${C}^{\rm perc}_{k k'} = (k'-1) P(k' \, \vert \, k)$, as pointed out in
Ref.~\cite{morenopercolation}. It is easy to check that if $v_k^{\rm perc}$
is an eigenvector of $\mathbf{C}^{\rm perc}$ with eigenvalue $\Lambda$,
then $\tilde{v}_k = k \; v_k^{\rm perc}$ is an eigenvector of
$\mathbf{\tilde{C}}$ with the same eigenvalue. Then, the eigenvalues
of $\mathbf{C}^{\rm perc}$ and $\mathbf{\tilde{C}}$ coincide, yielding
in this way the same description and an identical critical point.

\subsection{Correlated Scale-Free networks}

The discussion of the absence of epidemic threshold of the SIS in SF
networks with any sort of degree correlations can be easily extended
to the SIR model, taking again advantage of the Frobenius theorem. In
this case, in the general inequality given by Eq.~(\ref{eq:10}), we
set $\mathbf{A} = \mathbf{\tilde{C}}^2$ and $\psi(k)=k$, obtaining
\begin{equation}
  \tilde{\Lambda}_m^2\geq \min_k \sum_{k'} \sum_{\ell} (\ell-1) P(\ell  \, \vert \, k) (k' -1)
  P(k'  \, \vert \, \ell ).
\end{equation}
Given that $\sum_{k'} (k' -1)  P(k'  \, \vert \, \ell ) =
\bar{k}_{nn}(\ell,k_c) -1$, the previous inequality reads
\begin{equation}
  \tilde{\Lambda}_m^2\geq \min_k \sum_{\ell} (\ell-1) P(\ell  \, \vert \, k) \left[
    \bar{k}_{nn}(\ell,k_c) -1 \right]. 
\label{eq:11}
\end{equation}
As in the case of the SIS model, the divergence of the ANND with $k_c$
in the thermodynamic limit, ensures the divergence of the eigenvalue
$\tilde{\Lambda}_m$. Therefore, a SF degree distribution with diverging
second moment is a sufficient condition for the absence of an epidemic
threshold also for the SIR model if the minimum degree of the network
is $k_{min}\geq 2$. The only instance in which we can have an infinite
$\bar{k}_{nn}(\ell,k_c)$ with a finite eigenvalue $\tilde{\Lambda}_m$ is when
the divergence of $\bar{k}_{nn}$ is accumulated in the degree $k=1$
and results canceled by the term $\ell -1$ in Eq.~(\ref{eq:11}). This
situation happens when the SF behavior of the degree distribution is
just due to vertices with a single edge that form star-like structures
by connecting on a few central vertices. Explicit examples of this
situation are provided in
Refs.~\cite{morenostructured,morenopercolation}.

\section{Conclusions}

In this paper we have reviewed the analytical treatment of the
epidemic SIS and SIR models in complex networks at different levels of
approximation, corresponding to the different levels in which degree
correlations can be taken into account. At the zero-th level, in which
all the vertices are assumed to have the same degree (homogeneous
networks), we observe the presence of an epidemic threshold,
separating an active or endemic phase from an inactive or healthy
phase, that is inversely proportional to the average degree $\avk$. At
this level of approximation, both models render the same result, thus
showing a high degree of universality. At the first order
approximation level, in which vertices are allowed to have a different
degree, drawn from a specified degree distribution $P(k)$, but are
otherwise random, we obtain epidemic thresholds that are inversely
proportional to the degree fluctuations $\fluck$. The remarkable fact
about this result is that the epidemic threshold vanishes for SF
networks with characteristic exponent $ 2 < \gamma \leq 3$ in the limit of
an infinitely large network. Finally, in the second order
approximation level, in which degree correlations are explicitly
controlled by the conditional probability $\condP$ that a vertex of
degree $k$ is connected to a vertex of degree $k'$, our analysis
yields that the epidemic threshold in the SIS and SIR models is
inversely proportional to the largest eigenvalue of the connectivity
matrices $C_{k k'} = k \condP$ and $\tilde{C}_{k k'} = k(k'-1) P(k' \,
\vert \, k) /k'$, respectively. In the case of the SIR model we
recover the mapping with percolation at the level of correlations
exclusively among nearest neighbor vertices. The analysis of the
divergence of the average nearest neighbors degree
$\bar{k}_{nn}(k,k_c)$ with the degree cut-off $k_c$ allows us to
establish the general result that any SF degree distribution with
diverging second moment is a sufficient condition for the vanishing of
the epidemic threshold in the SIS model. The same sufficient condition
holds in the SIR model with $k_{min}\geq 2$.  The SIR model with
$k_{min}=1$ always shows the absence of an epidemic threshold with the
exception of the peculiar case in which the divergence of the average
nearest neighbor degree is accumulated only on the nodes of minimum
degree. These results have extremely important consequences, since
they imply that correlations are not able to stop an epidemic outbreak
in SF networks, in opposition to previous claims, and indicates that a
reduction of epidemic incidence can only be obtained by means of
carefully crafted immunization strategies
\cite{psvpro,aidsbar,havlinimmuno}, or trivially through finite size
effects \cite{pvbrief}.

\section*{Acknowledgments}

This work has been partially supported by the European commission FET
Open project COSIN IST-2001-33555. R.P.-S. acknowledges financial
support from the Ministerio de Ciencia y Tecnolog{\'\i}a (Spain).

\end{document}